\documentstyle[aps,preprint]{revtex}

\newcommand{\be}{\begin{equation}}
\newcommand{\ee}{\end{equation}}
\newcommand{\bea}{\begin{eqnarray}}
\newcommand{\eea}{\end{eqnarray}}
\newcommand{\corr}[1]{\langle #1 \rangle}
\def\bp{{\bf p}}
\def\bn{{\bf n}}
\def\bq{{\bf q}}
\def\br{{\bf r}}
\def\bS{\hat{\bf S}}

\begin{document}

\tighten

\title{Weak antilocalization in a 2D electron gas with the chiral splitting
of the spectrum}
\author{M. A. Skvortsov}
\address{L. D. Landau Institute for Theoretical Physics, Moscow 117940, RUSSIA}
\date{\today}
\maketitle

\begin{abstract}
Motivated by the recent observation of the metal-insulator transition
in Si-MOSFETs
we consider the quantum interference correction to the conductivity
in the presence of the Rashba spin splitting.
For a small splitting, a crossover from the
localizing to antilocalizing regime is obtained.
The symplectic correction is revealed
in the limit of a large separation between the chiral branches.
The relevance of the chiral splitting for the 2D electron gas in Si-MOSFETs
is discussed.
\end{abstract}

\pacs{}

Since the appearance of the scaling theory of localization\cite{AALR} in 1979,
it was a common belief that there can be no metal-insulator transition (MIT)
in 2D electron systems since all the states are localized at arbitrary weak
disorder.
Recent experiments on high-mobility Si-MOSFETs
by Kravchenko et al\cite{Kravchenko&}
showed however an evidence for a MIT at zero magnetic field
which is controlled by the density $n_s$ of 2D carriers.
For small densities $n_s<n_c \simeq 10^{11} {\rm cm}^{-2}$ the system
is insulating with exponentially diverging resistivity in the limit $T\to0$,
whereas for $n_s$ higher than the critical density
a strong drop in resistivity (by one order of magnitude)
is observed for $T<2$K.

The origin of the new metallic phase has not been understood yet.
Nevertheless it is evident that the electron-electron interaction
plays an important role as the critical density, $n_c$, is quite low
so that the Coulomb interaction dominates the kinetic energy.
Their ratio is $r_s\simeq10$ at the transition point
and decreases $\propto n_s^{-1/2}$ deep into the metallic phase.
Several theoretical approaches to the treatment of the strong
Coulomb interaction such as $p$-wave\cite{Phillips2}, triplet\cite{Belitz}
or anyon\cite{Zhang} superconductivity and superconductivity
resulting from a negative dielectric function\cite{Phillips1}
were suggested during the last year.

Besides a strong Coulomb interaction Si-MOS structures are characterized
by a spin-orbit splitting of the spectrum\cite{spinsplit}.
It originates from a strong asymmetry of the confining potential $V(z)$
of the quantum well. The corresponding term in the Hamiltonian of 2DEG,
the so-called Rashba term, is given by\cite{BR}
\be
\label{Rterm}
  H_{SO} = \alpha [\hat\sigma\times\hat\bp].
\ee
Here $\hat\sigma$ is the vector of the Pauli matrices,
$\hat\bp$ is the 2D momentum operator,
$\alpha$ is a constant of the spin-orbit symmetry breaking
measured in the units of velocity,
and $[\cdot\times\cdot]$ stands for the $z$-component of the vector product.
This term lifts the spin-degeneracy at zero magnetic field
and results in the splitting of the spectrum into two chiral branches:
\be
\label{branches}
  \epsilon^{(0)}_{\pm}(p) = \frac{p^2}{2m} \pm \alpha p,
\ee
with the splitting growing linearly with $p$.

For a Si-MOSFET, the minimum of the spectrum (\ref{branches}),
$-\epsilon_0=-m\alpha^2/2$, is estimated as 0.3K \cite{PudalovP1}
while the Fermi energy is $\epsilon_F\simeq6$K at the transition.
Then the ratio of the concentrations of left- and right-chiral fermions is
$n_+/n_- = (\sqrt{\epsilon_F+\epsilon_0}+\sqrt{\epsilon_0})^2
/(\sqrt{\epsilon_F+\epsilon_0}-\sqrt{\epsilon_0})^2 \simeq 2.5$.
Thus we conclude that the spin splitting results in a drastic change
of the internal properties of the system even without allowing for the
Coulomb interaction. This observation may question the remark by
Belitz and Kirkpatrick\cite{Belitz} that the spin-orbit scattering
is irrelevant due to a long-ranged Coulomb interaction:
The latter should be strongly modified by the predominance
of one type of chirality.

The relevance of the spin correlations was also demonstrated
in magnetic measurements\cite{magnetic}. Magnetic field applied
in the 2D plane was shown to suppress the metallic state
leading to a huge increase in resistivity.
The measurements in a perpendicular magnetic field show
a large positive magnetoresistance at high densities $n_s>2n_c$
also indicating the spin-related origin of the conducting phase.

We argue that the understanding of the new conducting phase as well
as the MIT itself can hardly be obtained without taking the strong
chiral splitting into account. Thus the theory of the metallic state
should be the theory of the Coulomb interacting chiral fermions.
The necessary first step then is to consider the noninteracting particles
with the chiral splitting of the spectrum.

In this letter we study the first quantum correction to the conductivity
for the noninteracting particles
in the presence of the Rashba term (\ref{Rterm})
and obtain it as a function of the spin-orbit splitting.
There are three energy scales in the problem:
the first is the Fermi energy $\epsilon_F$,
the second is the chiral splitting $\Delta=2\alpha p_F$ between the two
branches (\ref{branches}) at the Fermi level,
and the third is the inverse elastic mean free time $\tau^{-1}$
introduced by disorder.
We will assume $\epsilon_F$ to be the largest energy scale:
\be
\label{EtD}
  \epsilon_F \gg \frac{1}{\tau}, \quad
  \epsilon_F \gg \Delta.
\ee
The relationship between $\Delta$ and $\tau^{-1}$ is not specified
so that the variable
\be
\label{x-def}
  x = \Delta\tau
\ee
that controls the strength of the chiral splitting
may vary from 0 to $\infty$
provided that the relations (\ref{EtD}) are fulfilled.
At the critical density, the ratio $\Delta/\epsilon_F$ is of the order of 1
but decreases as $n_s^{-1}$ into the metallic phase. The experimental
value of the parameter $x$ slightly depends on the density,
being of the order of 2 when $n_s$ varies from $10^{11}{\rm cm}^{-2}$
to $3\times10^{12}{\rm cm}^{-2}$.

The spin-orbit scattering {\em at random potential} is known to drive
the system into the symplectic ensemble resulting in an antilocalizing
correction to the conductivity
$\Delta\sigma_{\rm symp} = (e^2/\pi h) \ln(L_\varphi/l)$
\cite{Hikami&}, where $l$ is the mean free path,
$L_\varphi=(D\tau_\varphi)^{1/2}$ is the phase-breaking length
associated with the phase relaxation time $\tau_\varphi$, $D$ is the diffusion
coefficient. In the case of the Rashba term, SU(2) symmetry is broken
on the level of {\em the regular Hamiltonian} while the potential scattering
may be considered as spin independent.
We will show that the parameter $x$ controls the crossover between
the orthogonal and the symplectic universality classes:
the correction to the conductivity becomes antilocalizing
at $x_*=(l/L_\varphi)^{1/3}\ll1$ and approaches $\Delta\sigma_{\rm symp}$
for $x\geq1$ as one can anticipate from the symmetry consideration.

Weak localization effects
in the presence of different types of spin-orbit splittings,
including the Rashba one, were studied extensively in Refs.~\cite{Pikus&}.
However the authors were interested mainly in the behavior
of magnetoresistance while the quantum correction at zero magnetic field
and for $x\geq1$ when $H_{SO}$ cannot be treated as a small perturbation
had not been investigated.

We consider a 2D noninteracting electron gas with the Rashba term
in the Hamiltonian:
\be
\label{H}
  H = \frac{\hat\bp^2}{2m}
    + \alpha \hat{p}_y \sigma_x - \alpha \hat{p}_x \sigma_y + U(\br)
\ee
where $U(r)$ is a random spin-independent impurities' potential, which
for the sake of simplicity is assumed to be Gaussian
$\delta$-correlated: $\corr{U(\br)U(\br')}=\delta(\br-\br')/2\pi\nu\tau$.
Here $\nu=m/2\pi$ is the density of states
for the free Hamiltonian $\bp^2/2m$.

The classical conductivity can easily be shown to be independent on $x$
and given by the Drude formula $\sigma_0=ne^2\tau/m$
provided that the random potential is $\delta$-correlated.
The first quantum correction to the conductivity\cite{1qc} is given
by the expression\cite{comment1}
\be
\label{1corr}
  \Delta\sigma = - \frac{e^2}{h} \frac{v_F^2}{2}
  \int \frac{d^2\bp}{(2\pi)^2}
    \corr{G^R(\bp)}^{\rho\alpha} \corr{G^R(-\bp)}^{\lambda\sigma}
    \corr{G^A(-\bp)}^{\sigma\beta} \corr{G^A(\bp)}^{\mu\rho}
  \int_{1/L_\varphi}^{1/l} \frac{d^2\bq}{(2\pi)^2}
    C^{\alpha\lambda}_{\beta\mu}(\bq),
\ee
where $\corr{G^{R,A}}$ are disorder-averaged retarded (advanced)
Green functions which for our problem are nondiagonal in the spin space
and the static Cooperon propagator $C(\bq)$ is determined by the
ladder equation
\be
\label{C}
  C^{\alpha\lambda}_{\beta\mu}(\bq)
  = \frac{\delta^{\alpha\lambda} \delta^{\beta\mu}}{2\pi\nu\tau}
  + \frac{1}{2\pi\nu\tau} \int \frac{d^2\bp}{(2\pi)^2}
    \corr{G^R(\bp+\frac{\bq}{2})}^{\alpha\alpha'}
    \corr{G^A(-\bp+\frac{\bq}{2})}^{\beta\beta'}
      C^{\alpha'\lambda}_{\beta'\mu}(\bq).
\ee

The averaged Green function obeys the Dyson equation
$\corr{G(\bp)}^{-1} = G^{(0)}(\bp)^{-1} - \Sigma$,
where $G^{(0)}(\bp)$ is the Green function of the unperturbed Hamiltonian.
In the quasiclassical limit, $\epsilon_F\tau\gg1$,
only diagrams without intersections of impurity lines are important
and the self-energy function
$
  \Sigma_{R,A}^{\alpha\beta} = \frac{1}{2\pi\nu\tau}
  \int \frac{d^2\bp}{(2\pi)^2} \corr{G^{R,A}(\bp)}^{\alpha\beta}
$.
On solving the Dyson equation we obtain the Green function that
can be written near the poles as ($\bn=\bp/p$)
\be
\label{G}
  \corr{G^{R,A}(\bp)} =
  \frac{ - \xi(p) \pm \frac{i}{2\tau}
       + \frac{\Delta}{2} ( n_y \sigma_x - n_x \sigma_y ) }
     { \left(
         - \xi(p) - \frac{\Delta}{2} \pm \frac{i}{2\tau}
       \right)
       \left(
         - \xi(p) + \frac{\Delta}{2} \pm \frac{i}{2\tau}
     \right) }.
\ee
Here we have taken an advantage of $\Delta\ll\epsilon_F$ and
substituted $\alpha p$ by $\Delta/2$.
The relaxation times for the two chiral branches appear to be equal
to each other and coincide with the mean free time $\tau$.
This is a consequence of the model with $\delta$-correlated disorder.
For a more realistic model with finite correlation length the lifetimes
will be different for the two chiralities but the difference will
be small in the limit $\Delta\ll \epsilon_F$.

The crucial quantity that determines the spin structure of the Cooperon is the
integral of the retarded and advanced Green functions,
\be
  I^{\alpha\alpha'}_{\beta\beta'}(\bq)
  = \frac{1}{2\pi\nu\tau} \int \frac{d^2\bp}{(2\pi)^2}
    \corr{G^R(\bp+\frac{\bq}{2})}^{\alpha\alpha'}
    \corr{G^A(-\bp+\frac{\bq}{2})}^{\beta\beta'}.
\ee
Calculating this integral as a function of $x$,
expanding to the second order in $ql$,
and substituting into Eq.~(\ref{C}) we get
\be
\label{CA}
  \hat C(\bq) = \frac{\hat A^{-1}(\bq)}{2\pi\nu\tau},
\ee
where the operator $\hat{A}(\bq) = \hat{\openone} - \hat{I}(\bq)$
expressed in terms of the total Cooperon spin
$\bS = \frac12 (\hat\sigma^R + \hat\sigma^A)$ reads
\bea
  \hat{A}(\bq) = &&
  \frac12 q^2 l^2
  + x^2 \left( \frac1{2(1+x^2)} - \frac{6+3x^2+x^4}{8(1+x^2)^3} q^2 l^2 \right)
    (\bS^2 - \hat{S}_z^2)
\nonumber
\\
\label{AS}
  &&{}- \frac{x^2(6+3x^2+x^4)}{4(1+x^2)^3} (\bq \times \bS)^2 l^2
  - \frac{x}{(1+x^2)^2} (\bq \times \bS) l.
\eea

The next step is to invert the matrix $\hat{A}$ and to obtain the Cooperon.
According to Eq.~(\ref{AS}), the singlet mode is gapless while the
triplet sector acquires a gap proportional to $x$.
To study the lifting of the triplet sector consider first the case
of small $x\ll1$. Then, for $ql\gg x$, the spin structure of $\hat{A}$
may be neglected so that $\hat{A}^{-1} = \frac{2}{q^2l^2}\hat{\openone}$.
For $ql\leq x$, the triplet sector of the inverse operator $\hat{A}^{-1}$
becomes complicated,
with different triplet modes having different gaps
because of the low symmetry of Eq.~(\ref{AS}),
but this region does not contribute to the logarithmic integral over $q$.
So we may write
\be
\label{A^-1}
  \hat{A}^{-1} \simeq
    \frac{2}{q^2l^2} \left( 1 - \frac{\bS^2}2 \right)
    + \frac{2}{q^2l^2+x^2} \frac{\bS^2}2.
\ee
This is not an exact formula but it captures correctly log-large terms
in $q$-integration.

Inserting (\ref{A^-1}) to Eq.~(\ref{CA}) and performing the integration,
we obtain the expression for the Cooperon integral:
\be
\label{Cindex}
  \int_{1/L_\varphi}^{1/l} \frac{d^2\bq}{(2\pi)^2}
    C^{\alpha\lambda}_{\beta\mu}(\bq)
  = \frac{1}{8\pi^2\nu v_F^2\tau^3}
  \left\{
    \left( \ln\frac{L_\varphi}{l} + 3 f \right)
      \delta^{\alpha\lambda}\delta^{\beta\mu}
    - \left( \ln\frac{L_\varphi}{l} - f \right)
      \sum_{i=1}^3 \sigma_i^{\alpha\lambda} \sigma_i^{\beta\mu}
  \right\},
\ee
where the contribution of the triplet sector,
\be
\label{f}
  f \left( x,\frac{L_\varphi}{l} \right) =
  \cases{
    \ln\frac{L_\varphi}{l} & for $x\ll\frac{l}{L_\varphi}$; \cr
    \ln\frac{1}{x} & for $\frac{l}{L_\varphi}\ll x\ll1$; \cr
    O(1) & for $x\gg1 \vphantom{\bigr]}$.
  }
\ee

The last thing to do it to compute the integral of four Green functions
in Eq.~(\ref{1corr}):
\bea
  \int \frac{d^2\bp}{(2\pi)^2}
    \corr{G^A(\bp)}^{\mu\rho} \corr{G^R(\bp)}^{\rho\alpha}
    \corr{G^R(-\bp)}^{\lambda\sigma} \corr{G^A(-\bp)}^{\sigma\beta}
  =
\nonumber
\\
  \frac{4\pi\nu\tau^3}{1+x^2}
  \left[
    \left( 1+\frac{x^2}2 \right) \delta^{\mu\alpha} \delta^{\lambda\beta}
    - \frac{x^2}4
      \left(
        \sigma_x^{\mu\alpha} \sigma_x^{\lambda\beta} +
        \sigma_y^{\mu\alpha} \sigma_y^{\lambda\beta}
      \right)
  \right].
\label{Kindex}
\eea
This integral is diagonal in the spin space for small $x\ll1$
but has a more complex structure for $x\gg1$ when the chiral branches
are well separated.

Finally, we combine all together.
Substituting (\ref{Cindex}) and (\ref{Kindex}) into Eq.~(\ref{1corr}),
after some arythmetics with the Pauli matrices we obtain the final expression
\be
\label{answer}
  \Delta\sigma = \frac1\pi \frac{e^2}{h}
  \left\{
    \ln \frac{L_\varphi}l
    - \frac{3 + x^2}{1 + x^2} f \left( x, \frac{L_\varphi}l \right)
  \right\}.
\ee

Let us study $\Delta\sigma$ as a function of $x$ for a given
$L_\varphi \gg l$. For $x\ll l/L_\varphi$, the spin splitting can be neglected
and we obtain the orthogonal universality class correction
$\Delta\sigma_{\rm orth}$ which can be interpreted as a sum of
a localizing contribution from the triplet sector and
an antilocalizing contribution from the singlet sector.
Then, for $l/L_\varphi \ll x$, the triplet modes acquire a gap
that reduces their contribution and the total correction
changes its sign and becomes antilocalizing at
\be
  x_* = \left( \frac{l}{L_\varphi} \right)^{1/3}.
\ee
For $x\gg x_*$, the antilocalization becomes more pronounced,
approaching $\Delta\sigma_{\rm symp}$ at $x\geq1$.
The gapless singlet contribution appears to be $x$-independent
while the contribution of the triplet channel acquires a gap and is
additionally suppressed by the factor $\frac{3+x^2}{1+x^2}$
which results from the special structure of the integral (\ref{Kindex})
of four Green functions.

Summarizing, we present the behavior of $\Delta\sigma$ in the form
\be
  \Delta\sigma = \frac1\pi \frac{e^2}{h} \times
  \cases{
    - 2 \ln\frac{L_\varphi}{l} & for $x\ll\frac{l}{L_\varphi}$; \cr
    \ln \left( \frac{L_\varphi}l x^3 \right)
      & for $\frac{l}{L_\varphi}\ll x\ll1$; \cr
    \ln\frac{L_\varphi}{l} & for $x\gg1 \vphantom{\Bigr]}$.
  }
\ee

The crossover from the orthogonal to the symplectic corrections
obtained for $x\ll1$ is related to the appearance of the gap in the
triplet sector of the Cooperon.
Our result should be contrasted with the absence of the first quantum
correction obtained in Ref.~\cite{L-G} for a certain type of the spin-orbit
coupling.

In conclusion, we considered the quantum interference correction
to the conductivity of the noninteracting fermions in the presence
of the Rashba spin-orbit interaction.
At small chiral splittings, $x\sim x_*$, the correction changes
the sign and becomes antilocalizing.
It approaches the symplectic universal value $\Delta\sigma_{\rm symp}$
when the scattering between
the different chiralities is strongly suppressed.
The present theory may be considered as the step toward the
understanding of the conducting phase in Si-MOSFETs that are likely
made of the Coulomb-interacting chiral fermions.

It is a pleasure for me to thank V. M. Pudalov for the idea of this work
and M. V. Feigel'man for useful discussions.
I am greatly indebted to I. V. Gornyi, A. P. Dmitriev and V. Yu.\ Kachorovskii
for pointing out the mistake in the first version of this paper.
I acknowledge that this material is based upon
work supported by U.S. Civilian Research and Developement
Foundation (CRDF) under Award \# RP1-273,
INTAS-RFBR grant \# 95-0302, and Swiss National Science Foundation
collaboration grant \# 7SUP J048531.


\begin{references}
\bibitem{comment1}
In the presence of the Bychkov-Rashba term,
velocity is no longer proportional to
momentum: $\hat v_i = \hat p_i/m + \alpha \epsilon_{ij} \sigma_j$.
This should modify Eq.~(\ref{1corr}) but the corrections are small
in the limit $\Delta/\epsilon_F\ll1$.
\end{references}
\end{document}